\documentclass[aps,prl,reprint,twocolumn,superscriptaddress,floatfix,nofootinbib,longbibliography]{revtex4-1}
\usepackage{graphicx,amsmath,amsfonts,amssymb,amsthm,xr}
\usepackage{epsfig,amsmath,amssymb,color,dsfont,upgreek,physics}
\usepackage{mathrsfs}
\usepackage{widetable}
\usepackage{mathtools}
\usepackage{bbold}
\usepackage{comment}
\usepackage{float}

\usepackage[bookmarks=true,colorlinks,linkcolor=OrangeRed,urlcolor=NavyBlue,citecolor=RoyalBlue]{hyperref}
\usepackage[dvipsnames]{xcolor}
\usepackage{orcidlink}
\usepackage{pifont} 

\AtBeginDocument{\let\oldcontentsline\contentsline}
\newcommand{\notoccontentsline}[4]{\oldcontentsline{}{}{}{}}
\newcommand{\droptocpage}{\addtocontents{toc}{\let\protect\contentsline\protect\notoccontentsline}}
\newcommand{\incltocpage}{\addtocontents{toc}{\let\protect\contentsline\protect\oldcontentsline}}

\usepackage{lineno}
\begin{document}

\preprint{APS/123-QED}

\title{String breaking mechanism in a lattice Schwinger model simulator}

\author{Ying Liu}
\thanks{Y.L.,~and W.-Y.Z. contributed equally to this work.}
\affiliation{Hefei National Research Center for Physical Sciences at the Microscale and School of Physical Sciences, University of Science and Technology of China, Hefei 230026, China}
\affiliation{CAS Center for Excellence in Quantum Information and Quantum Physics, University of Science and Technology of China, Hefei 230026, China}

\author{Wei-Yong Zhang}%
\thanks{Y.L.,~and W.-Y.Z. contributed equally to this work.}
\affiliation{Hefei National Research Center for Physical Sciences at the Microscale and School of Physical Sciences, University of Science and Technology of China, Hefei 230026, China}
\affiliation{CAS Center for Excellence in Quantum Information and Quantum Physics, University of Science and Technology of China, Hefei 230026, China}

\author{Zi-Hang Zhu}%
\affiliation{Hefei National Research Center for Physical Sciences at the Microscale and School of Physical Sciences, University of Science and Technology of China, Hefei 230026, China}
\affiliation{CAS Center for Excellence in Quantum Information and Quantum Physics, University of Science and Technology of China, Hefei 230026, China}

\author{Ming-Gen He}%
\affiliation{Hefei National Research Center for Physical Sciences at the Microscale and School of Physical Sciences, University of Science and Technology of China, Hefei 230026, China}
\affiliation{CAS Center for Excellence in Quantum Information and Quantum Physics, University of Science and Technology of China, Hefei 230026, China}

\author{Zhen-Sheng Yuan}%
\affiliation{Hefei National Research Center for Physical Sciences at the Microscale and School of Physical Sciences, University of Science and Technology of China, Hefei 230026, China}
\affiliation{CAS Center for Excellence in Quantum Information and Quantum Physics, University of Science and Technology of China, Hefei 230026, China}
\affiliation{Hefei National Laboratory, University of Science and Technology of China, Hefei 230088, China}

\author{Jian-Wei Pan}%
\affiliation{Hefei National Research Center for Physical Sciences at the Microscale and School of Physical Sciences, University of Science and Technology of China, Hefei 230026, China}
\affiliation{CAS Center for Excellence in Quantum Information and Quantum Physics, University of Science and Technology of China, Hefei 230026, China}
\affiliation{Hefei National Laboratory, University of Science and Technology of China, Hefei 230088, China}

\date{\today}

\begin{abstract}
String breaking is a fundamental concept in gauge theories, describing the decay of a flux string connecting two charges through the production of particle-antiparticle pairs. 
This phenomenon is particularly important in particle physics, notably in Quantum Chromodynamics, and plays a crucial role in condensed matter physics.  
However, achieving a theoretical understanding of this non-perturbative effect is challenging, as conventional numerical approaches often fall short and require substantial computational resources. 
On the experimental side, studying these effects necessitates advanced setups, such as high-energy colliders, which makes direct observation difficult.
Here, we report an experimental investigation of the string breaking mechanism in a one-dimensional U(1) lattice gauge theory using an optical lattice quantum simulator. 
By deterministically preparing initial states of varying lengths with fixed charges at each end, and adiabatically tuning the mass and string tension, we observed \emph{in situ} microscopic confined phases that exhibit either string or broken-string states. 
Further analysis reveals that string breaking occurs under a resonance condition, leading to the creation of new particle-antiparticle pairs. 
These findings offer compelling evidence of string breaking and provide valuable insights into the intricate dynamics of lattice gauge theories. 
Our work underscores the potential of optical lattices as controllable quantum simulators, enabling the exploration of complex gauge theories and their associated phenomena.

\end{abstract}

\maketitle
\droptocpage

\textbf{\textit{Introduction.---}}Non-perturbative phenomena in quantum field theories, such as string breaking, are fundamental yet challenging to study both theoretically and experimentally. 
In Quantum Chromodynamics (QCD), the potential between two static color charges increases linearly with distance, preventing quarks from existing in isolation \cite{wilson1974confinement}. 
When the separation exceeds a critical distance, the energy becomes sufficient to produce a light quark-antiquark pair, leading to string breaking and causing the potential to saturate at the energy of two static-light mesons. 
This phenomenon lies at the heart of gauge theories and is crucial for understanding confinement and deconfinement dynamics in QCD \cite{bali2005observation,buyens2016confinement}.
However, studying string breaking remains difficult across various approaches, including lattice QCD, due to the complex interplay between string and two-meson states. 
Conventional Monte Carlo methods struggle to fully resolve this phenomenon, underscoring the necessity for innovative theoretical approaches and experimental platforms to deepen our understanding of this fundamental process \cite{hebenstreit2013real,verdel2020real,kuhn2015non,gattringer2016approaches,Magnifico2020realtimedynamics,verdel2023dynamical}.

Quantum simulations offer a promising alternative for addressing these challenges~\cite{georgescu2014quantum,daley2022practical,bauer2023quantum}. 
Recent advancements in the various physical systems -- such as superconducting circuits~\cite{reed2012realization,arute2019quantum,wu2021strong}, trapped ions~\cite{tan2021domain,feng2023continuous,guo2024site}, neutral atom arrays~\cite{ebadi2021quantum,scholl2021quantum,bluvstein2022quantum,graham2022multi}, and optical lattices~\cite{zhang2023scalable,zheng2022efficiently,wienand2024emergence,young2024atomic} -- mark significant progress in illustrating practical quantum advantages and developing quantum error correction strategies~\cite{chiaverini2004realization}. 
Furthermore, these advancements facilitate the realization of highly controllable analog quantum simulators, which are essential for investigating complex dynamics that are generally difficult to address with classical approaches, particularly on larger scales. 
Recently, a series of works exploring low-dimensional lattice gauge theory (LGT) has emerged \cite{zohar2015quantum,pichler2016real,zohar2017digital,surace2020lattice,aidelsburger2022cold}, including the experimental demonstration of building blocks of 1D $\mathbb{Z}_2$ and U(1) LGT~\cite{martinez2016real,bernien2017probing,schweizer2019floquet,gorg2019realization,mil2020scalable}, medium-scale quantum simulations of Coleman's phase transition and thermalization \cite{yang2020observation,zhou2022thermalization}, and the observation of confinement in U(1) LGT \cite{zhang2023observation}. 
Despite significant progress, direct observation of string breaking remains elusive in analog quantum simulators.

\begin{figure}[htb!]
    \centering     %
    \includegraphics[width=0.48\textwidth]{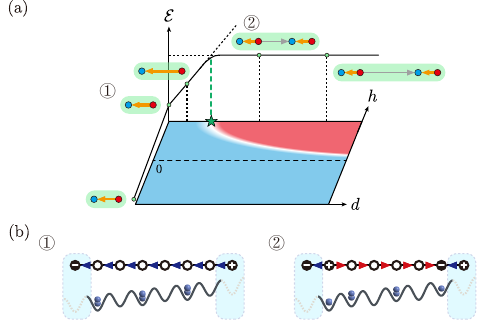}
    \caption{
    \textbf{String breaking in a lattice Schwinger model.}
    (a) Schematic illustration of the energy $\mathcal{E}$ between two static charges as a function of their separation distance $d$ and the string tension $h$. 
    In the large positive mass region of the U(1) lattice gauge theory, the energy $\mathcal{E}$ increases linearly with both the separation distance $d$ and the string tension $h$, until the strings break by creating particle pairs of mass $2m$, accompanied by a reversal in the orientation of $\mathcal{L} = d-2$ gauge sites. 
    (b) Configurations of the string state \ding{172} and the broken-string state \ding{173} in the Quantum Link Model (QLM, top) and the Bose-Hubbard Model (BHM, bottom) \cite{zhang2023observation,supp}. 
    For this illustration, the system size of the BHM is $L=6$, with a distance between two static charges of $d=L+1=7$ for the string state (left) and $\mathcal{L}=d-2=5$ flipped gauge sites for the broken-string state (right). 
    The light blue regions indicate the static electric field and charges.
    }
    \label{figure1:F1}
\end{figure}

In this work, we report the experimental observation of the string-breaking mechanism within the one-dimensional U(1) lattice gauge theory using an optical lattice quantum simulator. 
We establish a mapping of the target U(1) LGT with a tunable dynamical electric field, facilitated by a programmable optical superlattice complemented with a linear gradient tilt.
The optical superlattice provides parallel control capabilities that, combined with site-resolved addressing, allow for the deterministic preparation of initial states with varying lengths and two static charges at each end.
After adiabatically tuning the mass and string tension, we observe the prepared confined phases characterized by either string or broken-string states, distinguished by the production of particle-antiparticle pairs. 
Further analysis uncovers the resonance condition for string breaking, which occurs when the energy of the electric field equals the mass of the produced particle pairs. 
Our results provide compelling evidence of string breaking and significant insights into the non-perturbative aspects of lattice gauge theories.

\textbf{\textit{Model.---}}This work focuses on the study of a (1+1)-dimensional discretized form of quantum electrodynamics (QED), namely lattice Schwinger model, within the Quantum Link Model (QLM) framework~\cite{chandrasekharan1997quantum}, using a \textit{spin}-1/2 truncation. 
The Hamiltonian is expressed as follows:
\begin{equation}
\label{eq1:qlm}
\begin{aligned}
    \hat{H}_{\mathrm{QLM}} = & -\frac{\tilde{t}}{2} \sum^{}_{\ell}\left( \hat{\psi}_\ell^{\dagger} \hat{S}_{\ell, \ell+1}^{-} \hat{\psi}_{\ell+1} + \text{H.c.} \right) & \\
    & +m \sum_{\ell} (-1)^{\ell} \hat{\psi}_{\ell}^{\dagger} \hat{\psi}_{\ell} + h \sum_{\ell}^{} (-1)^{\ell+1} \hat{S}_{\ell, \ell+1}^z,
\end{aligned} 
\end{equation}
\noindent where, the fermionic operators $\hat{\psi}_{\ell}^{(\dagger)}$ denote the matter field with rest mass $m$ on site $\ell$. 
The operator $\hat{S}_{\ell,\ell+1}^{-}$ is the \textit{spin}-1/2 lowering operator acting on the link between sites $\ell$ and $\ell+1$, while $\hat{E}_{\ell,\ell+1} = (-1)^{\ell} \hat{S}_{\ell,\ell+1}^z$ represents the electric flux on the link.
The parameter $\tilde{t}$ defines the matter-gauge coupling strength, and $h$ represents the strength of the external background field, also referred to as string tension, associated with the topological $\theta$-angle term~\cite{halimeh2022tuning,cheng2022tunable,cheng2024emergent}. 

\begin{figure*}[htbp!]
    \centering %
    \includegraphics[width=1.00\textwidth]{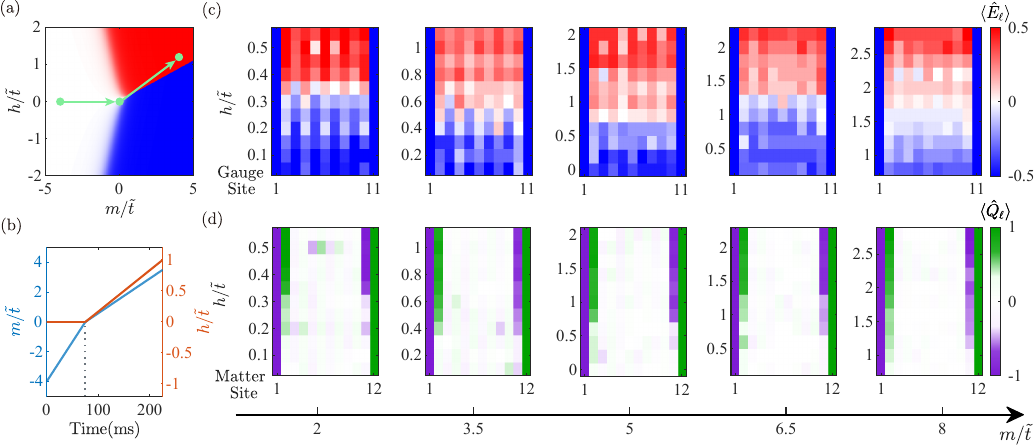}
    \caption{
    \textbf{Adiabatic preparation of string and broken-string states.}  
    (a) Numerically computed phase diagram for the QLM Hamiltonian (Eq.~\ref{eq1:qlm}) as a function of mass $m$ and string tension $h$, for a system size of $L=10$. 
    The calculated ground-state electric field $\langle E \rangle$ exhibits two distinct regions in the large positive mass condition: the string state (blue-shaded, with $\langle E \rangle < 0$) and the broken-string state (red-shaded, with $\langle E \rangle > 0$). 
    The green arrows represent the adiabatic parameter sweeps used for state preparation. 
    (b) Adiabatic ramp protocol for the mass parameter $m$ (light blue curve) and string tension $h$ (orange curve) during the experimental sequence.  
    (c) Experimentally measured electric field $\langle \hat{E}_\ell \rangle$ across gauge sites for a system with $L=10$, where $\langle \hat{E}_\ell \rangle$ simplifies to $\langle \hat{E}_{\ell,\ell+1} \rangle$. Subplots correspond to fixed $m$ values (left to right), with vertical axes showing $\langle \hat{E}_\ell \rangle$ variation with $h$.  
    (d) Experimentally measured charge $\langle \hat{Q}_\ell \rangle$ across matter sites for $L=10$, with subplots arranged similarly to (c).  
    }
    \label{figure2:F2}
\end{figure*}

\begin{figure*}[h]
    \centering
    \includegraphics[width=0.95\textwidth]{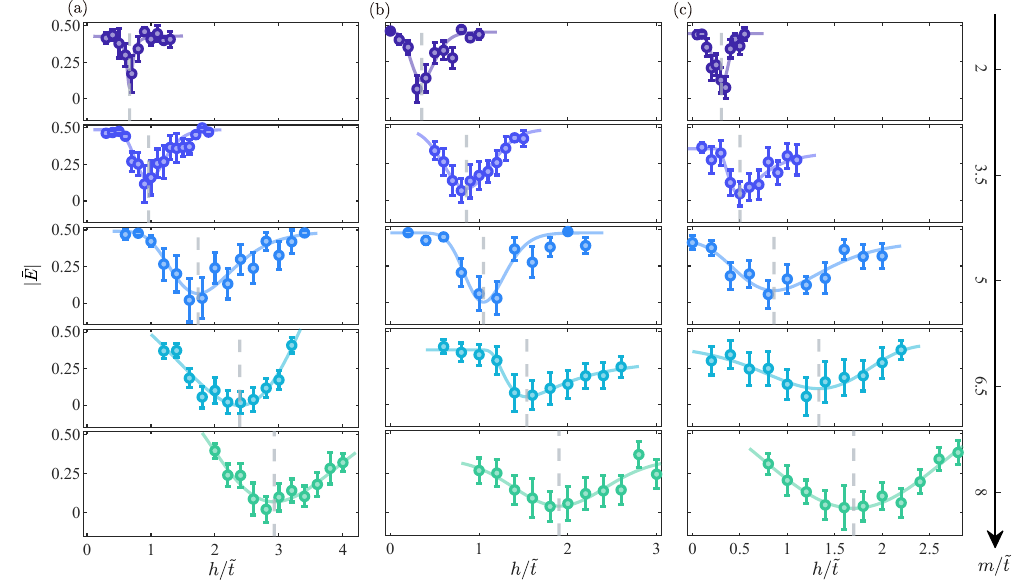}
    \caption{
    \textbf{Critical behavior for string breaking.}  
    Experimentally extracted absolute spatially averaged electric field as a function of string tension $h$ for the adiabatically prepared states at different rest masses $m$ (arranged from top to bottom) and system sizes: (a) $L=6$, (b) $L=8$, and (c) $L=10$. 
    The solid lines represent asymmetric Gaussian fits to the data points, and the dashed lines indicate the fitted position of the peak, corresponding to the critical value $h_c$ of the transition from string states to broken-string states.  
    }
    \label{figure3:F3}
\end{figure*}

\begin{figure*}[t!]
    \centering
    \includegraphics[width=1.00\textwidth]{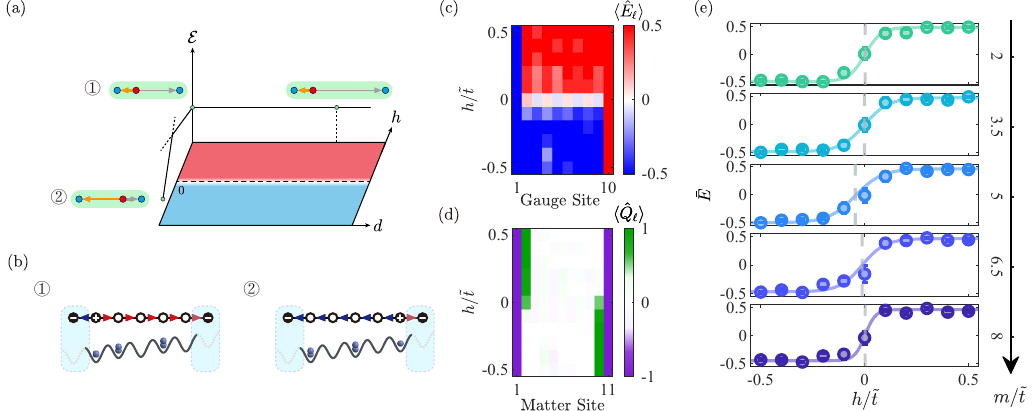}
    \caption{
    \textbf{Observation of string inversion.}  
    (a) Schematic plot of the energy $\mathcal{E}$ between two static charges of the same sign as a function of their separation distance $d$ and the string tension $h$ in the large positive mass regime. 
    (b) In the U(1) lattice gauge theory, Gauss's law enforces the presence of a dynamical electric charge with the opposite sign to the static charges for even $d$, forming a bound state with one of the static charges. 
    Consequently, all remaining electric field lines either point to the right (\ding{172}) or to the left (\ding{173}). 
    The energy $\mathcal{E}$ increases linearly with $h$ while remaining nearly independent of $d$.  
    Experimentally measured (c) electric field $\langle \hat{E}_\ell \rangle$ and (d) charge $\langle \hat{Q}_\ell \rangle$ for the final state in a system with $L=9$, showing their dependence on the string tension $h$.  
    (e) Spatially averaged electric field $ \bar{E} $ versus $h$ for different rest masses $m$ (top to bottom) in a system with $L=9$. 
    Solid lines represent hyperbolic tangent function fits, with dashed lines marking the critical value $h_c$ for string inversion.  
    }
    \label{figure4:F4}
\end{figure*}

\textbf{\textit{Experimental implementation.---}}To explore the string breaking mechanism, we designed experiments to observe the microscopic dynamics of string formation and breaking in the U(1) LGT. 
Fig.~\ref{figure1:F1}(a) provides a microscopic description of the string breaking process: a string connects two static charges of opposite signs accumulates energy linearly with either the string tension $h$ or the distance $d$ between the charges. 
When these parameters exceed a critical threshold, string breaking occurs, resulting in the creation of a new particle-antiparticle pair. 
This phenomenon is a manifestation of confinement, where the energy required to separate the charges leads to the production of new particles, thereby preventing the charges from existing independently.
In the experimental realization, directly observing the relationship between the energy of two static charges, their distance, and the string tension is difficult.
By preparing systems of various lengths and investigating their ground states under different string tensions, we can capture the essence of the string breaking process by identifying whether the system favors string or broken-string states. 

In our experiments, we implemented the target U(1) QLM (Eq.~\ref{eq1:qlm}) with ultracold $^{87}\mathrm{Rb}$ atoms in a programmable 1D optical superlattice~\cite{jaksch1998cold,greiner2002quantum}, incorporating a linear gradient tilt~\cite{sachdev2002mott,simon2011quantum,su2023observation}, as detailed in our previous works \cite{zhang2023observation,su2023observation} (see also the supplemental material \cite{supp}). 
We begin by preparing nearly defect-free arrays of $^{87}\mathrm{Rb}$ atoms, followed by initialization into a Fock state $ |1111 \ldots 1111\rangle $ with a deterministic length $L$ using the site-resolved addressing techniques \cite{zhang2023scalable}. 
The rest mass $m$ is initialized to a negative value by tuning the linear gradient tilt. 
Under the mapping relationship \cite{zhang2023observation,supp}, this Fock state corresponds to the ground state of the target U(1) QLM when $m \to -\infty$, with an oppositely charged particle-antiparticle pair at the two ends, provided the total number of sites is even. 

We then performed adiabatic evolution by gradually ramping the mass from negative to positive values, simultaneously adjusting the staggered potential to vary the final string tension $h$ across a range of values. 
This approach allowed us to prepare the system into various confinement phases, resulting in either a string or a broken-string state. 
The corresponding atomic configurations are $|2020\ldots 2020\rangle$ for the string state and $|1202\ldots 0201\rangle$ for the broken-string state, as illustrated in Fig.~\ref{figure1:F1}(b).
These final states were directly observed using our atom-number and site-resolved quantum gas microscope. 
To mitigate the effects of unwanted processes, such as atom loss, we applied two post-selection criteria: (i) conservation of total atom number, and (ii) conservation of Gauss’s law \cite{wang2023interrelated}. 

\textit{\textbf{Microscopic observation of string breaking.---}}We initially focus on a system with an even total number of sites, specifically with $L = 10$. 
The system consists of $L + 1$ gauge sites and $L + 2$ matter sites, with the two matter sites at each end fixed with static charges of opposite signs. 
Fig.~\ref{figure2:F2}(a) shows the electric field distribution of the ground state, calculated by averaging the electric fields of the $L - 1$ middle sites \cite{supp}.  
In the relatively large positive mass region, it is observed that as the string tension increases, the ground state of the system transforms from a string state to a broken-string state. 

For the experiment, we prepared a singly-occupied atom chain consisting of $L=10$ atoms, setting the mass to $m = -4 \tilde{t}$ and $h = 0$, achieving a substantial overlap with the ground state of the U(1) QLM system \cite{supp}. 
As shown in Fig.~\ref{figure2:F2}(b), we first adiabatically increased the rest mass to $m/\tilde{t} \approx 0$ in a linear fashion over 75 ms, with $\tilde{t} = 48.7(6)~\mathrm{Hz}$. 
Then, we adiabatically increased both the mass $m$ and string tension $h$ over a duration of 150 ms to various final values, driving the system into different confinement phases (refer to the supplementary materials \cite{supp} for more information on the adiabatic process). 
The final states were then read out using our quantum gas microscope.

Fig.~\ref{figure2:F2}(c) and (d) presents the extracted electric field, $\langle \hat{E}_{\ell,\ell+1} \rangle$, and the mean charge, $\langle \hat{Q}_{\ell} \rangle = \langle \hat{\psi}^\dagger_\ell \hat{\psi}_\ell +\big[(-1)^{\ell+1}-1\big]/2 \rangle$, as they change with the variation of string tension $h$, respectively, for different rest masses. 
These plots reveal that at smaller string tension $h$, the ground state tends towards a string state, characterized by the electric field pointing to the left across all sites in bulk. 
As the string tension increases, a new particle-antiparticle pair emerges, and the system transforms into a broken-string state, where the electric field points to the right across all sites in bulk. 
These results clearly indicate that the signature of string breaking can be identified when the string tension increases, while keeping the distance between charges fixed.

Additionally, we explored the impact of the distance between two static charges. 
This can be achieved by varying the length of the initial state. 
In our experiments, we prepared initial states with lengths of $L = 6$ and $L = 8$, respectively, and repeated the previously described experimental procedure. 
Fig.~\ref{figure3:F3} shows the absolute spatially averaged electric field $|\bar{E}| = | \frac{1}{L-1} \sum_{\ell=1}^{L-1} \langle \hat{E}_{\ell,\ell+1} \rangle |$ as a function of string tension for different rest masses and initial system sizes. 
To accurately determine the critical string tension at which the string breaks, we performed a modified asymmetric Gaussian fit on the spatially averaged electric field, using the position of the peak to represent the critical value $h_c$. 
The function is defined as $ A\times\exp[-\frac{1}{2}\frac{(x - h_c)^2}{(\sigma + b (x - h_c))^2}] + y_0$, where $ A,h_c,\sigma,b,y_0 $ are fitting parameters. 
The peak position $h_c$ is determined from the fit. 
The results from (a) to (c) in Fig.~\ref{figure3:F3} indicate that, for the same rest mass $m$, a longer string breaks at a smaller string tension. 

\begin{figure}[htbp]
    \centering     %
    \includegraphics[width=0.48\textwidth]{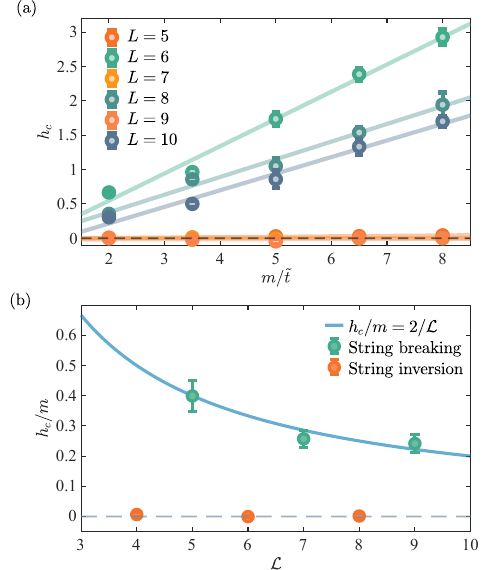}
    \caption{
    \textbf{Mechanism of string breaking and inversion.}  
    (a) Critical string tension $h_c$ extracted from the string breaking dynamics (Fig.~\ref{figure3:F3}) and string inversion process (Fig.~\ref{figure4:F4}) as a function of the rest mass $m/\tilde{t}$ for systems of varying sizes. 
    For string breaking dynamics, $h_c$ increases linearly with $m$, whereas for string inversion, $h_c$ remains near zero. 
    Solid lines indicate linear fits to the data.  
    (b) The slopes of the linear fits in (a) show that for string breaking dynamics, the ratio of the critical string tension $h_c$ to the rest mass $m$ is inversely proportional to the length of flipped gauge sites, $\mathcal{L}$. 
    This indicates that the resonance condition for string breaking is met when the energy of the flipped gauge sites equals the rest mass of the newly generated particle-antiparticle pair, i.e., $2m \approx h\mathcal{L}$.  
    In contrast, for string inversion, this ratio is nearly unaffected by changes in either the rest mass or string length.  
    }
    \label{figure5:F5}
\end{figure}

\textit{\textbf{Microscopic observation of string inversion.---}}Building upon our earlier discussions on string formation and breaking, we now delve into the intriguing phenomenon of ``string inversion'' in systems composed of an odd number of sites. 
According to Gauss’s law, two fixed static charges are required at both ends of the system. 
For systems with an odd number of sites, these static charges must possess the same sign, as dictated by the characteristics of the U(1) QLM Hamiltonian (Eq.~\ref{eq1:qlm}).  
At large positive mass values, as shown in Fig.~\ref{figure4:F4}(a), a dynamic charge inevitably emerges between the two static charges. 
This dynamic charge induces a reversal in the orientation of the string in the bulk, an effect we refer to as ``string inversion.'' 
Similarly, the energy map, as a function of system size and string tension shown in Fig.~\ref{figure4:F4}(a), highlights how string inversion manifests as the string tension increases. 

As the string tension shifts from negative to positive values, the dynamic charge, initially attracted to one of the static charges, migrates across the system to the static charge at the other end. 
This transition induces an inversion of the string in the bulk, thereby reorienting the connection between the static charges. 
Fig.~\ref{figure4:F4}(b) illustrates the corresponding ground state configuration in which the string is inverted, and the dynamic charge forms a pair with one of the static charges.  
This pairing effectively ``screens'' the other static charge, ensuring that the energy of the system remains invariant with respect to the distance between static charges. 

To experimentally investigate the string inversion phenomenon, we implemented systems with odd numbers of sites, specifically utilizing chain lengths of $L=5, 7, 9$.  
The adiabatic evolution procedure, as described earlier, was employed to prepare the ground state under varying rest masses and string tensions.  
Fig.~\ref{figure4:F4}(c) and (d) present the extracted electric field $\langle \hat{E}_{\ell,\ell+1} \rangle$ and mean charge $\langle \hat{Q}_\ell \rangle$ as a function of string tension $h$ for the $L=9$ system at a fixed positive rest mass.  
As the string tension varies, we observed distinct transitions consistent with the string inversion phenomenon.  
Additionally, Fig.~\ref{figure4:F4}(e) shows the spatially averaged electric field $\bar{E} = \frac{1}{L-1} \sum_{\ell=1}^{L-1} \langle\hat{E}_{\ell,\ell+1}\rangle$ depending on the string tension for different rest masses.  
Remarkably, the spatially averaged electric field undergoes an inversion near $h \sim 0$, largely independent of the rest mass. 
This result was further confirmed by fitting the data to a hyperbolic tangent function, with the fitted results (solid lines) showing a clear threshold at $h_c \sim 0$, indicating the onset of string inversion.

Moreover, as shown in Fig.~\ref{figure4:F4}(d), we find direct evidence for the dynamic charge transition across the chain as the string tension crosses zero. 
This charge transfer provides a definitive signature of string inversion within the system. 
Our analysis was extended to three odd-sized systems, $L=5, 7, 9$, yielding consistent results across all systems. 
Fig.~\ref{figure5:F5}(a) shows the extracted data for all three system sizes, demonstrating that string inversion is a general feature in odd-sized systems.  
The inversion consistently occurs when the string tension crosses zero, underscoring the robustness of this phenomenon across different system sizes.

\textit{\textbf{Resonance condition for string breaking dynamics.---}}To quantitatively analyze the string breaking mechanism, we turn our attention to the resonance conditions that govern the transition between string and broken-string states. 
By examining the fitting results shown in Fig.~\ref{figure3:F3}, we extract the relationship between the critical string tension $h_c$ at which string breaking occurs and the rest mass for different even-sized systems. 
The extracted data is presented in Fig.~\ref{figure5:F5}(a), where we observe a clear linear trend. 
This trend is substantiated by linear fitting (solid line), and the slope of this line provides valuable insights into the underlying dynamics of string breaking.

In Fig.~\ref{figure5:F5}(b), the slope derived from the fit is plotted, which reveals that the ratio of the rest mass to the string tension at the point of string breaking is inversely proportional to the number of flipped gauge sites, $\mathcal{L} = L-1$.
This result captures the resonance condition for string breaking: the energy of the flipped gauge sites at the transition point matches the energy of the particle-antiparticle pair created during the breaking process, i.e., $2m \approx h_c \mathcal{L}$. 
This result quantitatively explains the string breaking mechanism, highlighting that the transition occurs when the string tension balances the energy required to create the particle-antiparticle pair, minimizing the total energy of the system. 

\textbf{\textit{Conclusion and outlook.---}}In this work, we conducted a quantitative experimental study of the microscopic mechanism of string breaking in a U(1) lattice gauge theory using ultracold atoms in a programmable optical lattice. 
Our results distinctly demonstrate a transition from a string state to a broken-string state as string tension increases, accompanied by the emergence of particle-antiparticle pairs, thereby directly exemplifying the string breaking process. 
Additionally, we observed string inversion in odd-sized systems, where the dynamic charge presence induced a reversal of the string orientation when the string tension crossed zero. 
We also quantitatively assessed the resonance condition for string breaking, establishing a direct correlation among string tension, rest mass, and the energy required for particle-antiparticle pair creation.

Our experimental platform establishes a foundation for simulating more complex gauge theories and exploring non-perturbative phenomena in higher spin truncations \cite{osborne2305spin} and systems with higher spatial dimensions \cite{osborne2022large,felser2020two,magnifico2021lattice}.
Future research could extend these findings to investigate particle collisions\cite{surace2021scattering,rigobello2021entanglement,karpov2022spatiotemporal}, quantum criticality~\cite{wang2023interrelated}, localization-delocalization transition~\cite{karpov2021disorder,verdel2023dynamical}, and topological phases~\cite{wang2015topological,magnifico2019symmetry}.
Moreover, this work offers new avenues for quantum information science. 
Lattice gauge theories can be used to develop quantum error correction schemes \cite{carena2024quantum} and simulate high-energy physics models \cite{bauer2023quantum}, paving the way for novel quantum technologies.

\textbf{\textit{Note on related work.---}}
During the preparation of this manuscript, we became aware of related and complementary studies of string breaking dynamics with superconducting qubits \cite{cochran2024visualizing}, trapped ions \cite{de2024observation}, and Rydberg atoms \cite{gonzalez2024observation}.

\textbf{\textit{Acknowledgments.---}}This work was supported by NNSFC grant 12125409, Anhui Provincial Major Science and Technology Project 202103a13010005, Innovation Program for Quantum Science and Technology 2021ZD0302000. 
W.-Y.Z.~acknowledges support from the Postdoctoral Fellowship Program of CPSF under Grant Number GZC20241659.

\bibliography{main2arXiv}

\onecolumngrid
\vspace*{0.5cm}
\newpage
\begin{center}
    \textbf{METHODS AND SUPPLEMENTARY MATERIALS}
\end{center}
\vspace*{0.5cm}

\twocolumngrid
\incltocpage
\tableofcontents
\appendix
\setcounter{secnumdepth}{2}

\twocolumngrid
\setcounter{equation}{0}
\setcounter{figure}{0}
\makeatletter
\makeatother
\renewcommand{\theequation}{S\arabic{equation}}
\renewcommand{\figurename}{Extended Data Fig.}
\renewcommand{\thefigure}{\arabic{figure}}
\renewcommand{\thetable}{S\arabic{table}}

\section{Mapping between 1D U(1) lattice gauge theory and Bose-Hubbard model}

In this section, we provide a brief overview of the mapping between the (1+1)-dimensional U(1) Quantum Link Model (QLM) with \textit{spin}-1/2 truncation and the one-dimensional Bose-Hubbard Model (BHM) implemented in an optical superlattice. 
This mapping is essential for realizing the U(1) lattice gauge theory (LGT) dynamics using ultracold bosons, as discussed in the main text. 
More detailed descriptions can be found in our previous works and the references therein \cite{zhang2023observation,su2023observation,surace2020lattice}. 

\subsection{The U(1) Quantum Link Model}

The U(1) QLM Hamiltonian considered in this work is~\cite{chandrasekharan1997quantum}:
\begin{equation}
\label{eq:qlm_supp}
\begin{aligned}
    \hat{H}_{\mathrm{QLM}} = & -\frac{\tilde{t}}{2} \sum_{\ell}^{} \left( \hat{\psi}_\ell^{\dagger} \hat{S}_{\ell, \ell+1}^{-} \hat{\psi}_{\ell+1} + \text{H.c.} \right) \\
    & + m \sum_{\ell} (-1)^{\ell} \hat{\psi}_{\ell}^{\dagger} \hat{\psi}_{\ell} + h \sum_{\ell}^{} (-1)^{\ell+1} \hat{S}_{\ell, \ell+1}^z,
\end{aligned} 
\end{equation}
\noindent where, $\hat{\psi}_{\ell}^{(\dagger)}$ are fermionic annihilation (creation) operators representing matter fields at site $\ell$, and $\hat{S}_{\ell, \ell+1}^{-}$ is the spin-1/2 lowering operator acting on the link between sites $\ell$ and $\ell+1$. The operator $\hat{S}_{\ell, \ell+1}^z$ denotes the $z$-component of the spin operator, $\hat{E}_{\ell, \ell+1} = (-1)^{\ell} \hat{S}_{\ell, \ell+1}^z$ representing the electric flux on the link. The parameters $\tilde{t}$, $m$, and $h$ are the matter-gauge coupling strength, rest mass of the matter field, and the external background field (related to the topological $\theta$-angle term \cite{halimeh2022tuning,cheng2022tunable}), respectively.

\textbf{Gauss's law constraint.---}In this study, we ensure the local gauge invariance of the QLM through Gauss's law.
\begin{equation}
\label{eq:gauss_law}
\hat{G}_\ell = \hat{E}_{\ell,\ell+1}-\hat{E}_{\ell-1,\ell}-\hat{Q}_\ell = 0, \quad \forall \ell,
\end{equation}
\noindent where, $\hat{Q}_\ell=\hat{\psi}^\dagger_\ell \hat{\psi}_\ell +\big[(-1)^{\ell+1}-1\big]/2$. This constraint mandates that the physical states $\ket{\Psi}$ satisfy $\hat{G}_\ell \ket{\Psi} = 0$ for all sites $\ell$, meaning that the divergence of the electric field equals the matter charge at each site, consistent with Gauss's law in electromagnetism.

\subsection{The Bose-Hubbard model in a 1D optical superlattice}

The BHM used to simulate the QLM is given by:
\begin{equation}
\label{eq:bhm_supp}
\begin{aligned}
    \hat{H}_{\mathrm{BHM}} = & -J \sum_{\ell}^{} \left( \hat{b}_\ell^{\dagger} \hat{b}_{\ell+1} + \text{H.c.} \right) + \frac{U}{2} \sum_{\ell} \hat{n}_\ell (\hat{n}_\ell - 1) \\ 
    & + \sum_{\ell} \left[ \ell \Delta + (-1)^\ell \frac{\delta}{2} \right] \hat{n}_\ell.
\end{aligned} 
\end{equation}
In this Hamiltonian, $\hat{b}_\ell^{(\dagger)}$ are bosonic annihilation (creation) operators at site $\ell$, and $\hat{n}_\ell = \hat{b}_\ell^{\dagger} \hat{b}_\ell$ is the number operator. The parameters, $J$, $U$, $\Delta$, $\delta$, represent the hopping amplitude between neighboring sites, the on-site interaction strength, the linear gradient tilt potential between adjacent sites, and the staggered superlattice potential, respectively.

\textbf{Regime of interest.---}In this work, we focus on the parameter regime where $U \approx \Delta \gg J$, which energetically constrains the dynamics to a subset of states. In this limit, double occupancy is penalized by $U$, and the linear tilt $\Delta$ suppresses certain hopping processes due to the energy cost associated with moving against the tilt.

\subsection{Mapping procedure}

The mapping between the QLM and the BHM involves identifying the correspondence between the degrees of freedom and ensuring that the dynamics of the BHM reproduce the gauge-invariant dynamics of the QLM.

Since we are focusing on the near-resonant condition $U \approx \Delta \gg J$, the primary hopping process allowed is $\ket{1}_\ell \ket{1}_{\ell+1} \leftrightarrow \ket{2}_\ell \ket{0}_{\ell+1}$ between adjacent sites $\ell$ and $\ell+1$. 
As a result, only specific types of two-site configurations—(20, 11, 12, 02, 01)—are permitted at adjacent sites. Conversely, all other two-site configurations—(22, 21, 10, 00)—are forbidden under this near-resonance condition.

\textbf{Gauge field.---}The gauge field degrees of freedom are encoded in the bosonic occupations on adjacent sites $\ell$ and $\ell+1$. Two key configurations are:
\begin{itemize}
    \item $\ket{2}_\ell \ket{0}_{\ell+1}$: Corresponds to one configuration of the link.
    \item $\{ \ket{1}_\ell \ket{1}_{\ell+1}, \ket{1}_\ell \ket{2}_{\ell+1}, \ket{0}_\ell \ket{2}_{\ell+1}, \ket{0}_\ell \ket{1}_{\ell+1} \}$: Corresponds to the other configuration.
\end{itemize}

Therefore, we define an effective spin-1/2 degree of freedom on each link:
\begin{equation}
\begin{aligned}
    \ket{\uparrow}_{\ell,\ell+1} & \leftrightarrow \ket{2}_\ell \ket{0}_{\ell+1}, \\
    \ket{\downarrow}_{\ell,\ell+1} & \leftrightarrow  \ket{1}_\ell \ket{1}_{\ell+1}, \ket{1}_\ell \ket{2}_{\ell+1}, \ket{0}_\ell \ket{2}_{\ell+1}, \ket{0}_\ell \ket{1}_{\ell+1}.
\end{aligned}
\end{equation}

The spin operators acting on the link between sites $\ell$ and $\ell+1$ are defined as:

\begin{itemize}
    \item Raising operator:
    \begin{equation}
    \hat{S}_{\ell, \ell+1}^{+} = \ket{\uparrow}_{\ell, \ell+1} \bra{\downarrow}_{\ell, \ell+1}.
    \end{equation}
    \item Lowering operator:
    \begin{equation}
    \hat{S}_{\ell, \ell+1}^{-} = \ket{\downarrow}_{\ell, \ell+1} \bra{\uparrow}_{\ell, \ell+1}.
    \end{equation}
    \item $z$-component:
    \begin{equation}
    \hat{S}_{\ell, \ell+1}^{z} = \frac{1}{2} \left( \ket{\uparrow}_{\ell, \ell+1} \bra{\uparrow}_{\ell, \ell+1} - \ket{\downarrow}_{\ell, \ell+1} \bra{\downarrow}_{\ell, \ell+1} \right).
    \end{equation}
\end{itemize}

Thus, the electric flux on the link is given by $\hat{E}_{\ell, \ell+1} = (-1)^{\ell} \hat{S}_{\ell, \ell+1}^{z}$.

\vspace{6pt}

\textbf{Matter field.---}The matter field in the U(1) QLM is represented by fermionic matter fields, denoted by $\hat{\psi}_{\ell}^{(\dagger)}$, which are associated with each site $\ell$ of the 1D lattice. 
These fermions are coupled to the gauge field, which is encoded in the spin-1/2 operators on the links between adjacent sites. 
The fermionic matter field couples to the gauge field through the interaction terms in the Hamiltonian, which describe the hopping of fermions between sites and their interactions with the electric field on the links.

As we have already defined the gauge field in the U(1) QLM as mentioned above, then the matter field can be naturally determined by enforcing the Gauss's law constraint. 

\vspace{6pt}

Therefore, under the above mapping, the BHM Hamiltonian (Eq.~\ref{eq:bhm_supp}) effectively reproduces the QLM Hamiltonian (Eq.~\ref{eq:qlm_supp}) within the constrained Hilbert space.

\section{Experimental procedure}

\subsection{Initial state preparation}

Our experiments start with a two-dimensional (2D) Bose-Einstein condensate (BEC) of $^{87}\rm{Rb}$ atoms in the $5\mathrm{S}_{1/2} |F=1, m_{F}=-1\rangle$ state, confined to a single plane as described in our previous work~\cite{zhang2023scalable}. 
Staggered-immersion cooling~\cite{yang2020cooling} is then applied to produce a series of one-dimensional (1D) near-unity Mott insulators with a filling factor of $99.2(2)\%$.

Using site-resolved addressing techniques, we prepare multiple copies of atom chains with deterministic lengths, represented by the Fock state $|1111 \ldots 1111\rangle$.
According to the mapping between the U(1) QLM and the BHM, as detailed in the previous section, this Fock state $|1111 \ldots 1111\rangle$ in the BHM corresponds to the ground state of the U(1) QLM under the conditions $m \to -\infty$ and $h = 0$. This ground state represents a configuration of alternately arranged particles and antiparticles.
For finite negative rest mass, specifically at $m/\tilde{t} = -4$ in our experiment, the Fock state $|1111 \ldots 1111\rangle$ in the BHM also maintains a significant overlap with the ground state of the U(1) QLM. Numerical calculations confirm that this overlap exceeds $ 98\% $~\cite{zhu2024probingfalsevacuumdecay}.

\begin{figure*}[htb!]
    \centering     %
    \includegraphics[width=1.0\textwidth]{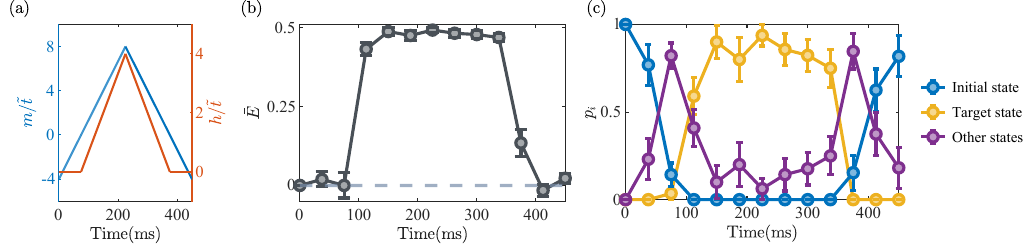}
    \caption{
    \textbf{Adiabaticity verification.}  
    (a) A ``round-trip'' ramp protocol for the rest mass $m$ (light blue) and string tension $h$ (orange) used to verify the adiabatic condition. 
    The forward ramp consists of two stages: the first stage costs 75 ms, and the second stage lasts 150 ms, as detailed in the main text and this supplementary material. 
    The backward ramp is the reverse of the forward ramp process.  
    (b) Real-time evolution of the spatially averaged electric field $\bar{E}$ during the round-trip ramp. 
    (c) Time evolution of the population distribution among the initial state (blue), target state (yellow), and other gauge-invariant states (purple). 
    }
    \label{figureS1}
\end{figure*}

\subsection{Adiabatic ramp protocol}

In our experiment, the ramp process is carefully designed to ensure the system evolves adiabatically, enabling a controlled study of the string breaking mechanism in the U(1) lattice gauge theory. The detailed ramping procedure and adiabaticity verification are described below. 

\textbf{Adiabatic ramp process.---} After preparing the initial Fock state $|1111 \ldots 1111\rangle$, a magnetic gradient tilt is applied to set the rest mass $m$ to a finite negative value, $m/\tilde{t} = -4$, with parameters $\tilde{t} = 48.7(6)~\mathrm{Hz}$, $U = 820(4)~\mathrm{Hz}$, and $\Delta = 430(2)~\mathrm{Hz}$. 
Details on the calibration of the relevant experimental parameters ($J$, $U$, $\Delta$, and $\delta$) can be found in our previous works \cite{zhang2023scalable,wang2023interrelated,zhang2023observation}.

Next, the lattice potential depth is rapidly quenched within 1 ms, bringing the system to its initial experimental conditions. 
Following this, the rest mass $m$ and string tension $h$ are adiabatically ramped in two stages. 
In the first stage, the rest mass $m$ is ramped from its initial value to nearly zero over 75 ms, while maintaining the string tension $h=0$. 
In the second stage, both the rest mass $m$ and string tension $h$ are simultaneously and adiabatically ramped over 150 ms to reach their respective target values, depending on the specific experimental requirements.

The entire ramp protocol is carefully designed to balance the adiabatic condition with experimental timescales, minimizing excitations to undesired states while ensuring a sufficient duration for observation.

\textbf{Adiabaticity verification through round-trip ramps.---} 
To verify whether the ramp process satisfies the adiabatic condition, we employed a ``round-trip'' ramp protocol. 
In this protocol, the system is first ramped to the target state at a controlled speed, as detailed in the previous section, and then ramped back to the initial state in reverse. 
The parameters of the entire ramp process are shown in Fig.~\ref{figureS1}(a). If the ramp process is adiabatic, the system should return to its initial state with high fidelity.

We implemented this test in an odd-sized $L=9$ system and set the target state parameters to $m/\tilde{t} = 8.0$ and $h/\tilde{t} = 4.0$, representing the largest variation in Hamiltonian parameters applied during the experiment. 
These values provide a stringent test of adiabaticity due to the significant changes in the energy landscape of the system. 
Fig.~\ref{figureS1}(b) shows the spatially averaged electric field $\bar{E}$ throughout the round-trip ramp process for this $L=9$ system. 
In the forward ramp, the spatially averaged electric field $\bar{E}$ is experimentally measured as $0.49(1)$ at the target state, showing excellent agreement with the theoretical prediction of 0.5. 
During the backward ramp process, as the system is ramped back to the initial experimental conditions, the spatially averaged electric field is measured to be $0.02(2)$, which is also in close agreement with the theoretical expectation of 0.

To further confirm the adiabaticity, Fig.~\ref{figureS1}(c) illustrates the real-time evolution of the probabilities for various states: initial state, target state, and other gauge-invariant states. 
The results indicate that while some other gauge-invariant states emerge during the middle of the ramp, the probability of the target state at the end of the forward ramp approaches 1 (measured as $ 93.8(6)\% $ experimentally). 
Similarly, at the end of the reverse ramp, the probability of the initial state also approaches a high value (measured as $ 82(1)\% $ experimentally). 
From these results, we estimate the fidelity of the target state at the end of the forward ramp to be approximately $ 90.4(3)\% $, demonstrating very high fidelity in the ramp process.

The observed high fidelity in the round-trip ramp process provides strong evidence that the system evolves adiabatically throughout the experiment. 
This ensures that the observed dynamics, such as the transition from the string state to the broken-string state, are intrinsic to the dynamics of the system and not artifacts caused by non-adiabatic excitations. 

\section{Numerical results}

To complement our experimental observations, we perform numerical simulations to investigate the phase diagram and critical properties of the U(1) QLM in the context of string breaking. 
The numerical results presented here and in the main text were obtained using exact diagonalization (ED) implemented with the Python package QuSpin~\cite{weinberg2019quspin}, providing deeper insights into the relation between string states and broken-string states, as well as the underlying resonance conditions. 

\begin{figure}[htb!]
    \centering     %
    \includegraphics[width=0.48\textwidth]{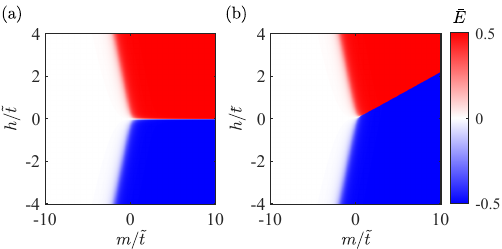}    
    \caption{
    \textbf{Phase diagram of the U(1) QLM.}
    Spatially averaged electric field $\bar{E}$ for a system with size: (a) $ L = 9 $, and (b) $ L = 10 $, where $L$ refers to the size of the BHM system. 
    }
    \label{figureS2}
\end{figure}

We numerically calculated the phase diagram of the U(1) QLM Hamiltonian (Eq.~\ref{eq:qlm_supp}) using ED and employed the spatially averaged electric field, defined as $\bar{E} = \frac{1}{L-1} \sum_{\ell=1}^{L-1} \langle\hat{E}_{\ell,\ell+1}\rangle$, as an order parameter. 
Below, we provide additional explanations for certain notations used consistently throughout the main text and this supplementary material. 
These clarifications aim to ensure precision and consistency in the description of the experimental and numerical results. 

\textbf{Explanation of notation and system size relation.---} In this work, $L$ represents the size of the physical system in the BHM, corresponding to the number of lattice sites used in the experiments. For a BHM system of size $L$, the associated U(1) QLM includes $L+1$ gauge sites and $L+2$ matter sites.
However, as noted in the main text, Gauss’s law fixes the charges at both ends of the QLM to static charges, and the gauge fields linked to these two static charges are also fixed.
As a result, our analysis focuses only on the middle $L-1$ gauge sites and $L$ matter sites. 
The order parameter $\bar{E}$ is defined as the average electric field over these $L-1$ gauge sites in the bulk of the system, excluding the fixed boundary fields. 
This definition is consistently applied throughout both the main text and this supplementary material.

\begin{figure}[htb!]
    \centering
    \includegraphics[width=0.48\textwidth]{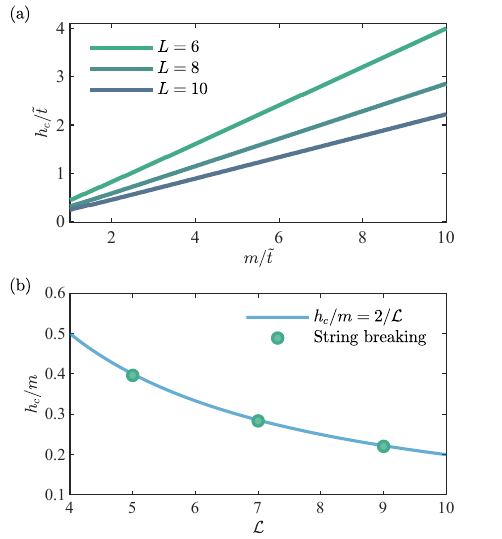}    
    \caption{
    \textbf{Microscopic mechanism of string breaking.}
    (a) Critical string tension $h_c$ as a function of mass $m$ for systems with even number of sizes $L$. The data are obtained from the sign-change in the spatially averaged electric field $\bar{E}$, marking the transition between string states and broken-string states.  
    (b) Slopes of $h_c$ versus $m$, showing an inverse proportionality to the string length $\mathcal{L}$. The results align with the resonance condition $2m \approx h_c \mathcal{L}$, confirming that the energy of the flipped gauge sites matches the energy required to generate a particle-antiparticle pair.  
    }
    \label{figureS3}
\end{figure}

\textbf{Phase diagram of the U(1) QLM.---} Fig.~\ref{figureS2}(a) and (b) show the spatially averaged electric field $\bar{E}$ for $L=9$ (odd-sized) and $L=10$ (even-sized) systems, respectively, where $L$ refers to the size of the BHM system. 

As discussed in the main text, when $L$ is odd, the static charges at the two ends of the system have the same sign. 
In the region of large positive mass, a dynamical charge emerges in the bulk and forms a bound state with one of the static charges. 
This bound state effectively ``screens'' the other static charge, causing the energy of the system to remain invariant with respect to the distance between the static charges. 
This unique behavior gives rise to a phenomenon referred to as ``string inversion.'' 
String inversion occurs when the sign of the string tension changes from negative to positive, or becomes inverted. 
As the string tension crosses zero, the dynamical charge shifts its bound state from one static charge to the other, causing the electric field in the bulk to reverse direction. 
This results in a reversal in the sign of the spatially averaged electric field $\bar{E}$. 
The point at which this sign change occurs marks the critical value of string tension, where the system undergoes string inversion. 

When $L$ is even, the static charges at the two ends of the system have opposite signs. 
In the large positive mass region, the system resides in a string state when the string tension is small, and the distance between the static charges is short.
As the string tension increases or the distance between the static charges grows, a pair of dynamical charges with opposite signs emerges in the bulk. 
This transition signifies the onset of the broken-string state. 
The transition from the string state to the broken-string state depends on both the string tension and the distance between the static charges. 
At the point of string breaking, the bulk electric field flips direction, causing the spatially averaged electric field $\bar{E}$ to change sign. 
The sign-change point indicates the critical string tension $h_c$ for string breaking, which depends on the system size and mass $m$.

\textbf{Critical behavior and resonance condition.---} To further investigate the microscopic mechanism of string breaking, we extracted the critical string tension $h_c$ as a function of mass $m$ for various even-sized systems, as shown in Fig.~\ref{figureS3}(a). 
Linear fits were applied to extract the slopes of the $h_c$-$m$ relationship for different system sizes. 

As shown in Fig.~\ref{figureS3}(b), for sufficiently large $m$, the slopes closely match $2/\mathcal{L}$, where $\mathcal{L}=L-1$ represents the number of flipped gauge sites during string breaking. 
This result demonstrates that the ratio $h_c/m$ is inversely proportional to $\mathcal{L}$, consistent with the resonance condition $2m \approx h_c \mathcal{L}$. 
This condition signifies that the energy of the flipped gauge sites matches the energy required to create a particle-antiparticle pair.

These numerical results support the experimental findings presented in the main text, further validating that our experiments have meaningfully contributed to uncovering the underlying mechanism of string breaking.

\end{document}